\documentclass[twocolumn,aps,prl]{revtex4}
\usepackage{graphicx}
\begin{document}
\title{Novel Orbital Ordering induced by Anisotropic Stress in a Manganite Thin Film}
\author{Y.~Wakabayashi$^1$, D.~Bizen$^2$, H.~Nakao$^2$,  Y.~Murakami$^{2,3}$, M.~Nakamura$^4$\footnote{Present address: Correlated Electron Research Center (CERC), National Institute of Advanced Industrial Science and Technology (AIST), Tsukuba, Ibaraki 305-8562, Japan}, Y.~Ogimoto$^5$, M.~Izumi$^6$, K.~Miyano$^6$ and H.~Sawa$^1$}
\address{$^1$Photon Factory, Institute of Materials Structure Science, High Energy Accelerator Research Organization, Tsukuba 305-0801, Japan\\
$^2$Department of Physics, Tohoku University, Sendai 980-8578, Japan\\
$^3$Synchrotron Radiation Research Center, JAERI, Sayo, 679-5148, Japan\\
$^4$Department of Applied Physics, University of Tokyo, Tokyo 113-8586, Japan\\
$^5$Devices Technology Research Laboratories, SHARP Corporation, Nara 632-8567, Japan\\
$^6$Research Center for Advanced Science and Technology, University of Tokyo, Tokyo 153-8904, Japan\\
}

\date{\today}
\begin{abstract}
We performed resonant 
and nonresonant x-ray diffraction studies of a Nd$_{0.5}$Sr$_{0.5}$MnO$_3$ 
thin film that exhibits a clear first-order transition. 
Lattice parameters vary drastically at the metal-insulator transition 
at 170~K ($=T_{\mbox{\scriptsize  MI}}$), and superlattice 
reflections appear below 140~K ($=T_{\mbox{\scriptsize  CO}}$). The 
electronic structure between $T_{\mbox{\scriptsize  MI}}$ and 
$T_{\mbox{\scriptsize  CO}}$ is identified as $A$-type antiferromagnetic 
with the $d_{x^2-y^2}$ ferroorbital ordering. 
Below $T_{\mbox{\scriptsize  CO}}$, a new type of antiferroorbital ordering
emerges. The accommodation of the large lattice distortion at the 
first-order phase 
transition and the appearance of the novel orbital ordering are 
brought about by the anisotropy in the substrate, a new parameter for the 
phase control.  
\end{abstract}

\maketitle

Charge ordering and orbital ordering ($CO$/$OO$) are the characteristic 
phenomena, which render the complex electronic phase behavior to the strongly
correlated electron systems, manganites in particular\cite{Tokukra00Science}.
A number of theoretical and experimental studies on $CO$/$OO$ in 
$RE_{1-x}AE_{x}$MnO$_3$ ($RE$: rare earth metals; $AE$: alkali earth metals) 
have been conducted in the vicinity of $x=0.5$ in order to understand the 
mechanism of the ordering and the resulting electronic properties
\cite{Fang00PRL}. Only three types of $OO$ have been found dominant --- 
one type of antiferroorbital structure (staggered arrangement of 
$d_{3x^2-r^2}$ and $d_{3y^2-r^2}$ orbitals, $CE$-$OO$) corresponding to 
$CE$-type antiferromagnetism ($AF$) and two types of ferroorbital structures 
($d_{3z^2-r^2}$ 
and $d_{x^2-y^2}$) corresponding to $C$-type and $A$-type $AF$, respectively. 

The orbital order couples intimately to the lattice distortion. 
One can easily envision that a tetragonal lattice distortion promotes 
ferroorbital structures; the compressive strain within the $c$-plane favors 
$d_{3z^2-r^2}$ ($C$-$OO$), while the tensile strain favors 
$d_{x^2-y^2}$ ($A$-$OO$). In fact, the phase control of ferroorbital ordering 
was achieved by manipulating the tetragonal lattice parameters employing 
a thin-film technique fabricated on (001) substrates\cite{Konishi99JPSJ}. 
In contrast, the antiferroorbital ordering inevitably involves the in-plane 
anisotropy and no effective means for its control has been available thus far. 

Thin manganite films on (011) substrates\cite{subst} were recently found to exhibit a variety of clear first-order phase transitions\cite{Ogimoto,Nakamura}, which has not been possible in those on (001) substrates that studied extensively\cite{Prellier00PRB,Ogimoto01APL,Biswas01PRB,Buzin01APL}.
 From the transport and magnetic properties of these films, the 
antiferroorbital order has been anticipated in them, although the direct 
evidence of the $OO$ as well as the knowledge of the $OO$ structures in 
these films, which affect the magnetic and/or electronic properties, were lacking. In this Letter, we present results of synchrotron x-ray diffraction measurements on a Nd$_{0.5}$Sr$_{0.5}$MnO$_3$ thin film grown on SrTiO$_3$ (011). A novel antiferroorbital structure has been identified. 
We clearly demonstrate a new handle to manipulate the $OO$, the {\it anisotropic} stress.

\begin{figure}
\includegraphics[width=7cm]{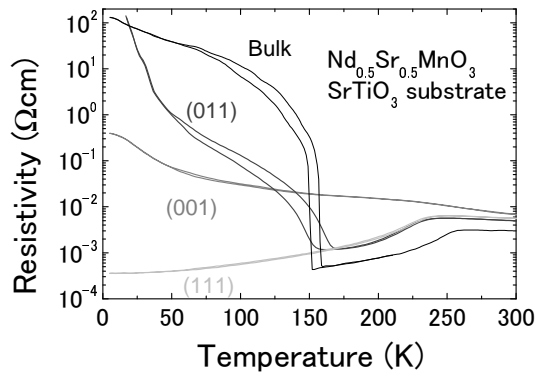}
\caption{Temperature dependence of the electric resistivity of Nd$_{0.5}$Sr$_{0.5}$MnO$_3$ thin films grown on SrTiO$_3$ (001), (011), and (111) substrates\protect{\cite{Nakamura}} and of bulk Nd$_{0.5}$Sr$_{0.5}$MnO$_3$\protect{\cite{Tokura96PRL}}.}
\label{fig:properties}
\end{figure}

The x-ray diffraction experiment was carried out at BL-4C and BL-16A2 of the 
Photon Factory, KEK, Japan. The beamlines are equipped with standard 
four-circle diffractometers connected to closed-cycle refrigerators. Epitaxial 
films were grown by the pulsed laser deposition method\cite{Nakamura,Ogimoto}. 
The thickness of the sample was 80~nm.  Figure \ref{fig:properties} shows the 
temperature dependence of the resistivity of Nd$_{0.5}$Sr$_{0.5}$MnO$_3$ thin 
films grown on SrTiO$_3$ (001), (011), and (111) substrates along with that of 
bulk Nd$_{0.5}$Sr$_{0.5}$MnO$_3$. The film on the 
(011) substrate clearly shows 
the first-order insulator-metal phase transition while films on other 
substrates show only monotonous temperature dependence. The transition,
however, is not as sharp as that in the bulk sample and the temperature
dependence of the resistivity is also different, which will be shown below to 
be a signature of the new $OO$ in the film. It should be noted that the 
electronic and magnetic properties show no essential thickness dependence\cite{Ogimoto}.

\begin{figure}
\includegraphics[width=7cm]{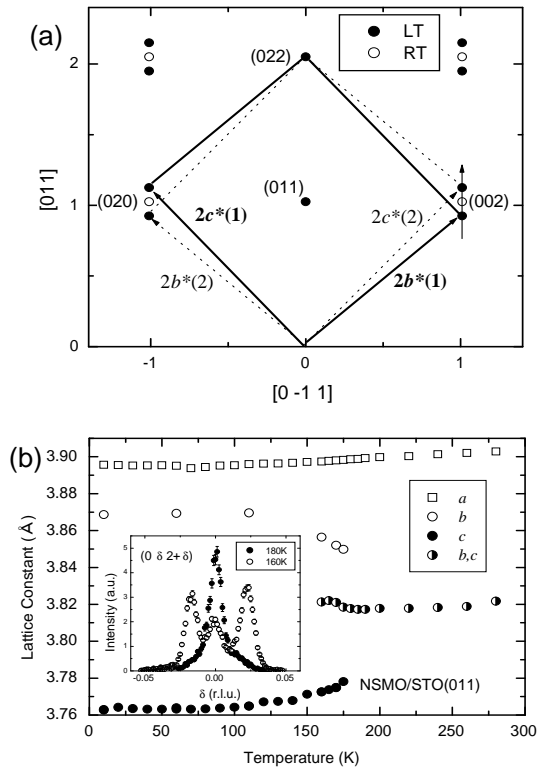}
\caption{(a) Schematic view of the $a$*-plane in reciprocal space at room temperature and 10~K. Twinning was observed at 10~K. (b) Temperature dependence of the lattice parameters $a$, $b$, and $c$. The lattice parameter of SrTiO$_3$ is shown as $a$ because $a$ was locked into the lattice parameter of the substrate that can be measured precisely. The inset shows the peak profiles obtained at 180~K and 160~K along the arrow through (002) shown in (a).}
\label{fig:recipro}
\end{figure}

First, we investigated the distortion in the primitive perovskite cell. A 
schematic view of the $a$*-plane in the reciprocal space is shown in 
Fig.~\ref{fig:recipro}(a). The lattice constants at room temperature are 
$a=3.905$~\AA, $b=c=3.824$~\AA, $\alpha=90.5^\circ$, and 
$\beta =\gamma =90.3^\circ$. As reported earlier\cite{Ogimoto,Nakamura}, 
the lattice constant $a$ is locked to the substrate, while that for 
[$0\bar11$] is unlocked. The (002) reflection splits into four peaks at 
10~K, i.e., ($\pm 0.008$, $+0.028$, $2+0.028$) and ($\pm 0.008$, $-0.028$, 
$2-0.028$). The closed circles in Fig.~\ref{fig:recipro}(a) show a schematic 
view of the reciprocal lattice at 10~K. The split along the $a$*-direction 
is ignored as it is very small. The lattice parameters at 10~K are 
$a=3.896$~\AA, $b=3.867$~\AA, $c=3.761$~\AA, $\alpha$=90.4$^\circ$, 
$\beta$=90.1$^\circ$, and $\gamma$=90.6$^\circ$. 
The temperature dependence of the lattice constants during a heating run is 
shown in Fig.~\ref{fig:recipro}(b). Lattice parameters $b$ and $c$ vary 
drastically at $T_{\mbox{\scriptsize  MI}}$, while lattice parameter $a$, 
which is locked to the substrate, is almost constant. 
This freedom of lattice parameters allows the first-order transition  
where the resistivity rapidly changes; epitaxial films on (001) substrates are tetragonally locked and this type of distortion is suppressed. The peak profiles of a 
(011) line passing through the (002) position at 160~K and 180~K are shown 
in the inset of the figure. The profile clearly shows the phase coexistence 
around 160~K.
The lattice parameters observed below $T_{\mbox{\scriptsize  MI}}$ are 
characterized by $a \simeq b > c$, similar to those observed in the orbital 
ordering in bulk manganites\cite{Kajimoto99PRB}. This feature is shared by 
$A$-$OO$ and $CE$-$OO$, which are commonly observed in manganites in the hole 
concentration of $x\simeq 0.5$\cite{Nakamura99PRB}. Therefore, the temperature 
dependence of the lattice parameters suggests that $A$-$OO$ or $CE$-$OO$ is 
established below $T_{\mbox{\scriptsize MI}}$.

Next, we searched for superlattice reflections corresponding to the orbital 
ordering. Superlattice reflections characterized by the wavevector ($\frac12 \frac12 \frac12$) are 
observed at room temperature. These reflections are caused by the 
MnO$_6$-octahedra rotation and the concomitant displacement of $A$-site 
(Nd and Sr) ions.  At 10~K, the intensity of these reflections differs from 
that observed at room temperature, indicating that the magnitude of the 
MnO$_6$-octahedra rotation changes with the phase transition. In addition, 
the superlattice reflections characterized by the wavevectors 
($\frac14 \frac14 \frac12$) and ($\frac12 \frac12 0$) were observed at this 
temperature. The size of the unit cell at 10~K is $\sqrt 2 \times 2\sqrt 
2\times 2$ times the primitive perovskite cell, which is the same as that 
of the bulk compounds exhibiting $CE$-$OO$. The wavevector of ($\frac12 \frac12 0$) is 
identical to that of the charge ordering in many bulk compounds.  

\begin{figure}
\includegraphics[width=7cm]{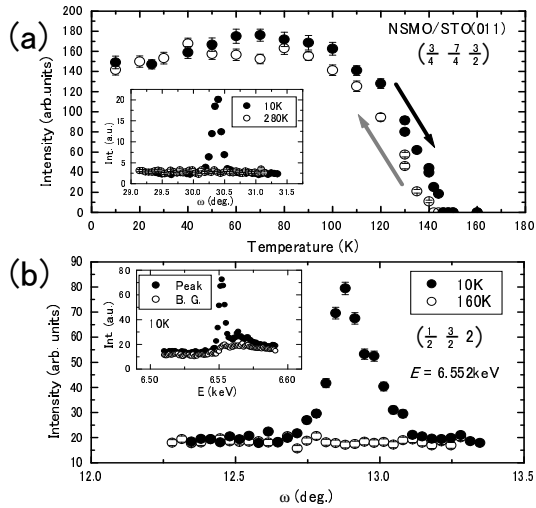}
\caption{(a) Temperature dependence of ($\frac34 \frac74 \frac32$) intensity measured using incident x-rays of 6.5~keV, {\it off-resonant} energy. The open and closed circles indicate the results of cooling and heating run, respectively. The inset shows the intensity distribution around ($\frac34 \frac74 \frac32$) at 10~K and 280~K. (b) Intensity distribution around ($\frac12 \frac32 2$) measured using incident x-rays of Mn $K$-edge energy at 10~K and 160~K. Inset shows the energy spectrum of this reflection at 10~K. }
\label{fig:super}
\end{figure}

The inset of Fig.~\ref{fig:super}(a) shows the intensity distribution around 
($\frac34 \frac74 \frac32$) at 10~K and 280~K. Clearly a new peak emerges at 
low temperatures. The intensity of this reflection as a function of 
temperature is shown in Fig.~\ref{fig:super}(a). The peak appears at 140~K 
($=T_{\mbox{\scriptsize  CO}}$) during the cooling run. This temperature is 
significantly lower than $T_{\mbox{\scriptsize  MI}}$ at which the (002) 
reflection splits. We searched for superlattice reflections at 160~K, the 
temperature between $T_{\mbox{\scriptsize  CO}}$ and 
$T_{\mbox{\scriptsize  MI}}$, and found no peak except for the 
($\frac12 \frac12 \frac12$) reflections. The electronic state in this 
temperature region will be discussed later.

Figure~\ref{fig:super}(b) shows the peak profile of the ($\frac12 \frac32 2$) 
reflection at 10~K, measured using the x-rays of Mn $K$-edge energy. The 
energy spectrum of the peak intensity at 10~K is shown in the inset. It is 
enhanced by a factor of 30 at the Mn $K$-edge. This indicates that two or more 
non-equivalent Mn sites form a periodic arrangement with the wavevector 
($\frac12 \frac12 0$), the ordinary charge ordering. This reflection was 
not observed above $T_{\mbox{\scriptsize  CO}}$, as shown in the figure. 
The temperature dependence of the intensity of this reflection is almost 
the same as that of the ($\frac34 \frac74 \frac32$) reflection.

\begin{figure}

\includegraphics[width=7cm]{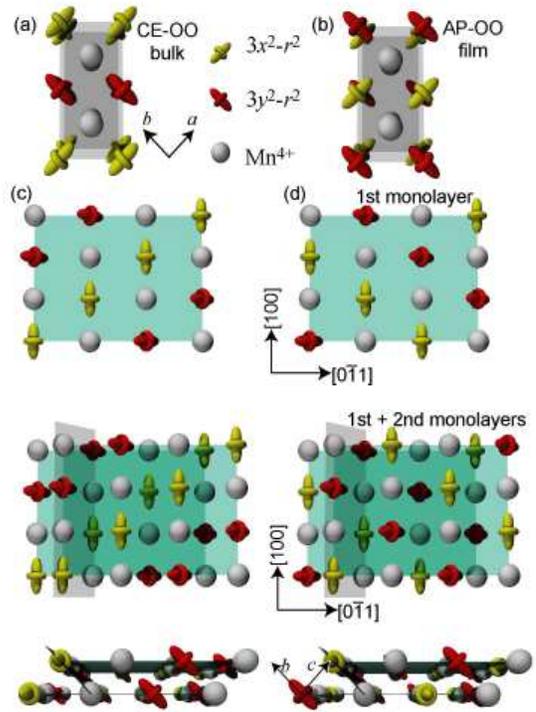}

\caption{(color) (a) Schematic view of the orbital arrangement in bulk compounds ($CE$-$OO$). The Mn$^{4+}$ ions are indicated by white spheres, while Mn$^{3+}$ ions with the $3x^2-r^2$ ($3y^2-r^2$) orbital are indicated by yellow (red) symbols. (b) The same as (a) but for the film ($AP$-$OO$). (c) The orbital arrangement of $CE$-$OO$ observed from the [011] direction (upper two figures) and [$\bar1 0 0$] direction (bottom figure). The [011] planes, which are parallel to the surface of the substrate, and the [001] planes are represented by green and gray transparent plates, respectively. (d) The same as (c) but for $AP$-$OO$.}
\label{fig:str}
\end{figure}

The superlattice reflections with the wavevector ($\frac14 \frac14 \frac12$) 
observed at 10~K imply that the in-plane structure is the same as 
that of $CE$-$OO$ with ($\frac14 \frac14 0$) diffraction. However, the 
stacking pattern alternates along the $c$-direction.
 The regular $CE$-$OO$ structure is shown in Fig.\ref{fig:str}(a). The 
in-plane orbital arrangement is common to systems having a variety of 
structure types, i.e., single-layered\cite{Murakami214}, 
bi-layered\cite{Argyriou00PRB}, and $A$-site ordered 
manganites\cite{Uchida02JPSJ,Arima02PRB,Kageyama03JPSJ}, while the stacking 
vectors are different. Therefore, it is reasonable to expect that our 
film has also the same in-plane orbital arrangement. 
Under this assumption, we arrive at a unique solution of the orbital 
arrangement in the film. The result is shown in Fig.~\ref{fig:str}(b). We 
term this arrangement antiphase-$OO$ ($AP$-$OO$). It is quite natural 
that the observed lattice 
parameters ($a\simeq b>c$) are similar to those found in the bulk $CE$-$OO$ 
and $A$-$OO$.

It should be noted that the $AP$-$OO$ must carry a magnetic 
structure different from that of $CE$-$OO$. The magnetic structure of the 
$AP$-$OO$ state in the $c$-plane must be the same as that of the $CE$-$AF$, 
i.e., the antiferromagnetic arrangement of zig-zag ferromagnetic chains 
resulting from the anisotropic ferromagnetic interaction of ordered $e_g$ 
orbitals. The stacking structure in the $CE$-$AF$ is antiferromagnetic; 
in contrast, perfect antiferromagnetic stacking is impossible in the 
$AP$-$OO$ because of the shift of the phase in the zig-zag chains in 
the neighboring planes. Under this constraint, two types of stacking 
structures are possible: the neighboring Mn$^{3+}$ spins in the $c$-direction 
are parallel and Mn$^{4+}$ spins antiparallel or vice versa. In both 
cases, lines of inter-plane transfer thread through the $c$-planes resulting
in a unique network structure. This structure is 
consistent with nearly isotropic but slightly enhanced electric conductivity 
in the [0$\bar1$1] direction\cite{Ogimoto}.

 The $AP$-$OO$ has not been observed in the bulk compounds thus far. 
The novel orbital arrangement is evidently stabilized by the strain 
from the substrate. For a given zig-zag pattern in the $c$-plane, the orbital 
arrangement within a [011] plane in $CE$-$OO$ and that in the 
$AP$-$OO$ are identical except for the rotation of 180$^\circ$ 
about the $a$-axis. This is shown in the upper 
sectors of Figs.\ref{fig:str} (c) and \ref{fig:str} (d), which indicate the first monolayers on the SrTiO$_3$ (011) substrates.
 Thus, the energy difference must come from the effect of the second monolayer.
 The second layer structures are shown in the middle and lower sectors of 
Figs.~\ref{fig:str} (c) and (d). For $CE$-$OO$, the stress exerted 
by the film on the substrate 
is non-uniform, because the locally 
distorted primitive perovskite cells stack up in phase. On the other hand, the 
stress in the $AP$-$OO$ is evenly distributed within 
the [011] plane since the arrangement is 
staggered. This difference can produce the energy gain of the $AP$-$OO$ 
arrangement. It should be stressed that a uniquely defined crystallographic 
axis, as is demonstrated here, is of great importance for the study of 
macroscopic anisotropic properties and can be rarely achieved in bulk single 
crystals\cite{Tobe}.

The stacking 
structure of the orbital ordering may also be 
affected by the distortion of the 
$A$-site ions through the hybridization of $A$-site ions and 
oxygens\cite{Mizokawa99PRB}. Since the unit cells of the film are strained 
by the substrate, the magnitude of the displacement of the $A$-site ions 
should differ from that of the bulk compound. This may change the stacking 
structure of the orbitals. However, detailed theoretical calculations of 
some orbital arrangements as a function of the $A$-site displacement has been 
carried out only for $x=0$\cite{Mizokawa99PRB}. Theoretical attention
 for $x=0.5$ 
is needed for detailed analysis.


Finally, we discuss the electronic state of the film between 
$T_{\mbox{\scriptsize  CO}}$ and $T_{\mbox{\scriptsize  MI}}$. 
As mentioned earlier, the lattice parameter $c$ is significantly 
smaller than the other two for this temperature range. This is the 
characteristic of $CE$-$OO$, $AP$-$OO$, and $A$-$OO$. However,
$CE$-$OO$ and $AP$-$OO$ should produce the superlattice 
reflections characterized by 
the wavevectors ($\frac14 \frac14 0$) and ($\frac14 \frac14 \frac12$), 
which are not observed in this temperature range. Therefore, the expected 
orbital state is $A$-$OO$. Since magnetism is closely related to the orbital 
state, the suppression of the spontaneous magnetization below $T_{\rm MI}$ 
(the $A$-type antiferromagnetic state)\cite{Ogimoto,Nakamura} 
is a natural consequence. 
The occurrence of $A$-type $AF$ in this film is not surprising 
because the free energy of the $A$-type $AF$ is similar to that of $CE$-$OO$, 
and thus that of $AP$-$OO$ for Nd$_{1-x}$Sr$_x$MnO$_3$ system. In 
fact, the bulk compound with $x=0.51$ exhibits the $A$-type $AF$ below 
200~K\cite{Kajimoto99PRB}. 

In summary, a new type of orbital ordering was observed in a thin film of 
Nd$_{0.5}$Sr$_{0.5}$MnO$_3$ fabricated on a SrTiO$_3$ (011) substrate. 
The orbital arrangement is clarified and the formation mechanism of this 
new orbital structure is discussed. The temperature dependence of the 
lattice parameters and the orbital ordered state are determined. $A$-type 
$AF$/$d_{x^2-y^2}$ ferroorbital order also appears at the intermediate 
temperature region. These results clearly show that various orbital structures 
can be realized in thin films, including the ones that do not appear in the
bulk crystal. The anisotropic stress from the substrate is a new parameter 
for controlling the electronic state in addition to the ionic radii and the 
hole 
concentration. 

The authors are grateful to Prof. T.~Arima and Dr. J.~P.~Hill for fruitful discussions.
This work was supported by a Grant-in-Aid for Creative Scientific Research (13NP0201)
and TOKUTEI (16076207) from the Ministry of Education, Culture, Sports, Science 
and Technology of Japan and JSPS KAKENHI (15104006). Financial support to M. N. 
by the 21st Century COE Program for ``Applied Physics on Strong Correlation'' 
administered by Department of Applied Physics, The University of Tokyo 
is also appreciated.


\begin{thebibliography}{99}
\bibitem{subst} Note that the notation of the crystallographic axes is different from earlier reports\cite{Ogimoto,Nakamura}.
\end{thebibliography}
\end{document}